\documentclass[
  prstab,
  superscriptaddress,
  twocolumn,%
  floatfix,%
  showkeys,
  final
]{revtex4}

\usepackage{amsmath}
\usepackage{amssymb}

\usepackage{graphicx}

\usepackage{enumerate}

\usepackage{fancyhdr}
\usepackage{upgreek}

\newcommand{\reportnum}{000000}

\fancypagestyle{first}{%
\fancyhf{}
\fancyfoot[C]{\reportnum-\thepage}
}

\renewcommand{\v}[1]{\vec{#1}}
\newcommand{\grad}{\nabla}
\newcommand{\density}{n}
\newcommand{\wbbeta}{\varphi}
\newcommand{\micron}{$\upmu$m}

\begin{document}

\title[Modulator simulations for coherent electron cooling \\ using a variable density electron beam]%
      {Modulator simulations for coherent electron cooling \\ using a variable density electron beam}
\thanks{This work is supported by the US DOE Office of Science, Office of Nuclear Physics,
grant numbers DE-SC0000835 and DE-FC02-07ER41499;
resources of NERSC were used under contract No. DE-AC02-05CH11231.}
\author{George I. Bell}
\email{gibell@txcorp.com}
\author{Ilya Pogorelov}
\affiliation{Tech-X Corporation, 5621 Arapahoe Ave, Suite A, Boulder CO 80303, USA}
\author{Brian T. Schwartz}
\affiliation{Lathrop and Gage, LLP, 4845 Pearl East Circle, Suite 201, Boulder CO, 80301, USA}
\author{David L. Bruhwiler}
\affiliation{Department of Physics, University of Colorado, 80309, USA}
\author{Vladimir Litvinenko}
\author{Gang Wang}
\author{Yue Hao}
\affiliation{Brookhaven National Lab, Upton NY 11973, USA}

\begin{abstract}
Increasing the luminosity of relativistic hadron beams is critical for
the advancement of nuclear physics. Coherent electron cooling (CEC) promises
to cool such beams significantly faster than alternative methods. We
present simulations of 40 GeV/nucleon Au+79 ions through the first (modulator) section
of a coherent electron cooler.
In the modulator, the electron beam copropagates with the ion beam, which
perturbs the electron beam density and velocity via anisotropic Debye
shielding.
In contrast to previous simulations, where the electron density was constant in time and space,
here the electron beam has a finite transverse extent, and undergoes focusing by quadrupoles
as it passes through the modulator.
The peak density in the modulator increases by a factor of $3$, as specified by the beam Twiss parameters.
The inherently 3D particle and field dynamics is modeled with the parallel
VSim framework using a $\delta$f PIC algorithm.
Physical parameters are taken from the CEC proof-of-principle experiment
under development at Brookhaven National Lab.
\end{abstract}


\maketitle 

\thispagestyle{first}

\section{Introduction}
Coherent electron cooling (CEC) is a novel technique for rapidly cooling high-energy, high-intensity
hadron beams \cite{Lit09}.
The proposed Brookhaven CEC consists of three sections:
a \textit{modulator}, where the ion imprints a density wake on the electron distribution,
a \textit{free electron laser} (FEL), where the density wake is amplified, and
a \textit{kicker}, where the amplified wake interacts with the ion,
resulting in dynamical friction for the ion.

In this paper we consider only the modulator section.
We simulate the wake in the electron distribution due to the presence of a single ion,
as the ion drifts with many co-propagating electrons.
Although the beam particles are highly relativistic in the laboratory frame,
particle velocities are non-relativistic in the beam frame drifting with 
the mean speed of the particles.
In previous work the electron density was
constant and uniform in space \cite{Bell11}, or the density was changing slowly in time \cite{Bell12}.
Here we consider a more realistic beam where the electron density decreases to zero at the edge of the beam,
and also changes in time, taking into account external focusing.

We assume the electron beam in the modulator is close to a known solution described
by the beam emittance and Twiss parameters.
The electron density in the beam changes in space as well as in time.
Simulation results are calculated using VSim (formerly Vorpal) \cite{Bell08} using
$\delta$f PIC \cite{NX06}.
The $\delta$f particles do not represent deviations of the beam from a steady-state, but rather
deviations of the beam away from the known solution described by Twiss parameters.

Wang and Blaskiewicz \cite{WB08} found exact solutions to the Vlasov-Poisson equations
in a uniform electron density assuming a special form of the electron velocity distribution,
a $\kappa$-2 distribution.
Despite the different assumptions of this model,
it is very easy to calculate,
so this simple model provides a useful comparison for our simulation results.

Qian et. al. \cite{Qian94} applied a $\delta$f model to study an intense
beam in a periodic quadrupole focusing channel.
They use a Kapchinskil-Vladimirskij (KV) distribution for their known solution
This distribution is particularly tractable analytically,
and allows them to include space charge effects.
Here we do not consider the beam as moving through a periodic focusing channel,
we consider only a single passage through the modulator.
We will also ignore space charge effects.

\section{Beam Frame and Lab Frame}
In general we work in the beam frame moving with the average speed of the particles.
However, beamline elements are at rest in the lab frame, and we will need to convert between the
lab and beam frames.
The direction of particle movement will be along the $z$-axis with speed $\beta c$,
where $\beta$ is just slightly smaller than $1$. 
For a particle in the beam frame with position $(x^{beam},y^{beam},z^{beam})$
and velocity $(v_x^{beam},v_y^{beam},v_z^{beam})$, its coordinates and
velocity in the lab frame are given by the Lorentz Transform:
\begin{eqnarray}
x^{lab} & = & x^{beam}, \label{eq:xconv} \\
y^{lab} & = & y^{beam}, \label{eq:yconv} \\
z^{lab} & = & \gamma z^{beam} + \gamma\beta c t^{beam}, \label{eq:zconv} \\
t^{lab} & = & \gamma\beta c^{-1} z^{beam} + \gamma t^{beam}, \label{eq:tconv} \\
v_x^{lab} & \approx & v_x^{beam}/\gamma, \label{eq:vxconv} \\
v_y^{lab} & \approx & v_y^{beam}/\gamma, \label{eq:vyconv} \\
v_z^{lab} & \approx & \beta c + v_z^{beam}/\gamma^2, \label{eq:vzconv}
\end{eqnarray}
where $\gamma = 1/\sqrt{1-\beta^2}$ is the Lorentz factor.
We assume in these formulas that all beam frame velocities are non-relativistic, or that
$|\v{v}^{beam}| \ll c$.
The velocity transformations ignore terms of order $|\v{v}^{beam}|/c$.
In what follows, we will usually work in the beam frame, and unless otherwise noted,
variables are in the beam frame.

\section{$\delta$f Formulation}
If $f(\v{x},\v{p},t)$ is the phase space electron density in the beam frame,
$f$ evolves according to the Vlasov equation, which
specifies that the total time derivative of $f(\v{x},\v{p},t)$ is zero,
\begin{equation}
\frac{Df}{Dt} = \frac{\partial f}{\partial t} + \frac{d\v{x}}{dt}\cdot\grad_x f + \frac{d\v{p}}{dt}\cdot\grad_p f = 0 .
\label{eq:vlasov}
\end{equation}
The particles accelerate due to the total electric field, which is
composed of the field $-\grad_x\phi(\v{x},t)$ due to the perturbing ion and electron charge distribution
plus an external (beam frame) electric field $\v{E}$ and magnetic field $\v{B}$,
\begin{equation}
\frac{d\v{p}}{dt} = e(-\grad_x\phi + \v{E} + \v{v}\times\v{B}) ,
\label{eq:momen}
\end{equation}
where $e<0$ is the electron charge.
The potential $\phi$ satisfies a self-consistent Poisson equation
\begin{equation}
\grad^2\phi = -\frac{\rho(\v{x},t)}{\epsilon_0} ,
\label{eq:poisson}
\end{equation}
where
\begin{equation}
\rho(\v{x},t) = Z|e|\delta(\v{x}-\v{x}_{ion})+e\tilde{n}(\v{x},t) ,
\label{eq:rho}
\end{equation}
$\tilde{n}(\v{x},t)=\int f(\v{x},\v{p}, t) d\v{p}$, and the ion located at $\v{x}_{ion}$ has charge $Z|e|$.

We now split the electron density
\begin{equation}
f=f^0+f^1 ,
\label{eq:split}
\end{equation}
where $f^0$ describes the bulk behavior of the beam,
and $f^1$ is a perturbation which describes the electron shielding response to the ion.
For the $\delta$f method the important thing is that
$f^0$ be a known (exact) solution to the Vlasov equation (\ref{eq:vlasov}).
For simplicity $f^0$ is often taken as an equilibrium solution.
Associated with $f^0$ is a potential $\phi_0$ which satisfies
a self consistent Poisson equation (\ref{eq:poisson}),
except that the ion is not present in Equation (\ref{eq:rho}).

The beam within the modulator is focused down by nearly a factor of two transversely
by a set of four quadrupoles.
We therefore chose a solution $f^0$ which is a function of the beam Twiss parameters.
This $f^0$ is not a steady-state solution,
but is a known solution to the Vlasov equation (\ref{eq:vlasov}).
This solution does not include the space charge term, which appears as the potential $\phi^0$
associated with the charge distribution $f^0$.

The perturbation density $f^1$ satisfies
\begin{equation}
\frac{Df^1}{Dt} = - e \grad_x (\phi^0 + \phi^1) \cdot \grad_p f^0 ,
\label{eq:wt0}
\end{equation}
where $\phi^1(\v{x},t)$ is the self-consistent potential for the perturbation $f^1$,
\begin{equation}
\grad^2\phi^1 = -\frac{\rho^1(\v{x},t)}{\epsilon_0} ,
\end{equation}
and
\begin{equation}
\rho^1(\v{x},t) = Z|e|\delta(\v{x}-\v{x}_{ion})+e\tilde{n}^1(\v{x},t) ,
\end{equation}
and $\tilde{n}^1(\v{x},t)=\int f^1(\v{x},\v{p},t) d\v{p}$.
We will calculate $\grad_p f^0$ analytically.
There are three contributions to the force on delta-f particles (\ref{eq:momen}).
First, the space charge on the beam, represented by the field $-\grad_x \phi^0$.
We assume space charge is negligible over the cooling section, so this term is dropped.
Second, the electric field due to the perturbation itself, 
$-\grad_x \phi^1$, and finally the external beam frame
fields $\v{E}$ and $\v{B}$.

The perturbation $f^1$ is modeled using $\delta$f PIC algorithm \cite{NX06}.
Suppose the $i$'th $\delta$f PIC particle has
position $\v{x}_i$, velocity $\v{v}_i$ and weight $w_i$.
These particles represent the perturbation $f^1$, as defined by
\begin{equation}
f^1(\v{x},\v{v},t) = \sum_i w_i \delta(\v{x}-\v{x}_i) \delta(\v{v}-\v{v}_i) .
\end{equation}
The initial weight of the $\delta$f particles is zero,
but their spatial and velocity distributions do not need to be the same as the background beam solution $f^0$.
The distribution of the $\delta$f particles we call $g$,
\begin{equation}
g(\v{x},\v{v},t) = \sum_i \delta(\v{x}-\v{x}_i) \delta(\v{v}-\v{v}_i) .
\end{equation}
It is important to choose a distribution function
$g$ which has sufficient resolution to capture the perturbation.
For an ion shielding perturbation which is localized in one region of space (around the ion),
it seems reasonable to choose a density function $g$ which is also localized in space.
We shall see that we also want to require that $Dg/Dt=0$ in order to
simplify the evolution of the particle weights.

The particle weights $w_i$ approximate a continuous weight function $w(\v{x},\v{v},t)$,
\begin{equation}
w(\v{x},\v{v},t) = f^1(\v{x},\v{v},t) / g(\v{x},\v{v},t)
\end{equation}
so that the total time derivative of $w$ is
\begin{equation}
\frac{Dw}{Dt} = \frac{1}{g} \left[ \frac{Df^1}{Dt} - \frac{f^1}{g} \frac{Dg}{Dt} \right]
\end{equation}
Assuming that we choose a distribution $g$ such that $Dg/Dt=0$, substituting (\ref{eq:wt0}) gives
\begin{equation}
\frac{Dw}{Dt} = - \frac{e}{g} \grad_x \phi^1 \cdot \grad_p f^0
\label{eq:wt1}
\end{equation}

Suppose first that the $\delta$f particles are distributed uniformly in phase space.
The $\delta$f particles move in response to the field $-\grad_x\phi_1+\v{E}$,
and their weights evolve according to the discrete version of Equation (\ref{eq:wt1}),
\begin{equation}
\frac{dw_i}{dt} = - \frac{e}{g} \grad_x \phi^1(\v{x}_i,\v{v}_i,t) \cdot \grad_p f^0 (\v{x}_i,\v{v}_i,t) ,
\label{eq:wt2}
\end{equation}
where $g$ is a constant equal to the phase space density of the $\delta$f particles.
For uniform loading in 6D phase space, between $x$-velocity bounds $v_x^{min}$ and $v_x^{max}$,
and similarly for $y$ and $z$,
\begin{equation}
g=\frac{\density}{(v_x^{max}-v_x^{min})(v_y^{max}-v_y^{min})(v_z^{max}-v_z^{min})}
\end{equation}

Alternatively, to initialize the $\delta$f particles we can use the distribution $f^0$.
Then $g=f^0+f^1$, $w = f^1/(f^0+f^1)$, and $1-w = f^0/g$,
so we can write our weight equation as
\begin{equation}
\frac{dw_i}{dt} = - e (1-w_i) \grad_x \phi^1(\v{x}_i,\v{v}_i,t) \cdot \frac{\grad_p f^0 (\v{x}_i,\v{v}_i,t)}{f^0 (\v{x}_i,\v{v}_i,t)}
\label{eq:wt3}
\end{equation}

A quadrupole of strength $q$ (Tesla/m) has a magnetic field in the lab frame
\begin{equation}
\v{B}^{lab} = q(y\hat{x} + x\hat{y}).
\label{eq:blab}
\end{equation}
Here we ignore the magnetic fringe fields, assuming that the magnetic field is
given by (\ref{eq:blab}) inside the quadrupole and is zero outside it.
If we convert the lab frame magnetic field (\ref{eq:blab}) into the beam frame
moving at velocity $\beta c\hat{z}$, we have
\begin{eqnarray}
\v{E}  & =  & \gamma\beta c q(- x\hat{x} + y\hat{y}), \label{eq:ebeam} \\
\v{B}  & =  & \gamma q(y\hat{x} + x\hat{y}).
\end{eqnarray}
In the beam frame particle velocities $\v{v}$ are non-relativistic,
so $|\v{v}\times\v{B}|\ll|\v{E}|$ and from here on we drop the term $\v{v}\times\v{B}$
from (\ref{eq:momen}).
The quadrupoles for the proof-of-principle experiment have strength $0.3$ KGauss/cm or
$q=3$ Tesla/m, a length (in the lab frame) of $16$ cm, and $\gamma=40$.
An electron which enters a quadrupole $1$ mm off center by (\ref{eq:ebeam}) experiences
a beam frame transverse accelerating gradient of $36$ MV/m.

\section{A beam defined by Twiss parameters}

In terms of Twiss parameters, the electron density $f^0$ in the lab frame
(see \cite{Rosenzweig}) is given by
\begin{equation}
f^0(\v{x},\v{x}') = f_x^0(x,x') f_y^0(y,y') f_z^0(z,v_z)
\label{eq:f0}
\end{equation}
where
\begin{equation}
f_x^0(x,x') = \frac{c_{\pi}}{\epsilon_x} \exp\left[- \frac{\gamma_x x^2 + 2\alpha_x xx' + \beta_x {x'}^2}{2\epsilon_x}\right]
\label{eq:f0x}
\end{equation}
and similarly for $f_y^0$,
\begin{equation}
f_y^0(y,y') = \frac{c_{\pi}}{\epsilon_x} \exp\left[- \frac{\gamma_y y^2 + 2\alpha_y yy' + \beta_y {y'}^2}{2\epsilon_y}\right]
\label{eq:f0y}
\end{equation}
Here $\alpha_x$, $\beta_x$, $\gamma_x$, $\alpha_y$, $\beta_y$ and $\gamma_y$ are standard Twiss parameters
which are functions of $z^{lab}$, and $c_\pi = 1/\sqrt{2\pi}$ is a normalization constant.
In the longitudinal direction ($z$), the distribution in velocity $v_z$ is Maxwellian about
the mean lab frame velocity $\beta c$.

\begin{figure}[htb]
\centering
\includegraphics[width=82mm]{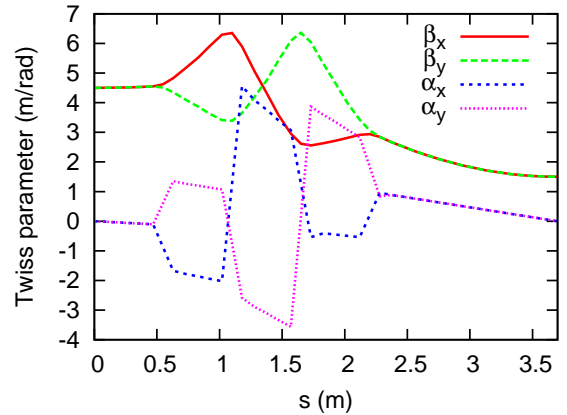}
\caption{The Twiss parameters $\beta_x$, $\alpha_x$, $\beta_y$ and $\alpha_y$ versus distance along the modulator.
$\beta_x$ and $\beta_y$ have dimensions of m/rad while $\alpha_x$ and $\alpha_y$ are dimensionless.}
\label{fig:1}
\end{figure}

Figure~\ref{fig:1} shows a plot of $\beta_x$, $\alpha_x$, $\beta_y$ and $\alpha_y$ through the modulator section (3.7 m in length).
The three Twiss parameters are related by the formula $\beta_x\gamma_x = 1 + \alpha_x^2$ (and similarly for $y$).
$\epsilon_x$ and $\epsilon_y$ are constants, the transverse rms emittance of the beam.
The longitudinal function $f_z^0(z,z')$ is uniform in $z$ and has a conventional Maxwellian velocity distribution.

Note that equation (\ref{eq:f0x}) specifies the density in the lab frame,
where $x'=v^{lab}_x/v^{lab}_z$ is the trajectory angle, assumed to be small.
We need to convert this density function to the beam frame.
The transverse coordinates $x$ and $y$ are the same in either coordinate system, from (\ref{eq:xconv}) and (\ref{eq:yconv}).
Using (\ref{eq:vxconv}) and (\ref{eq:vzconv}) we have
$x'\approx v^{beam}_x/(\gamma\beta c)$, since $v_z^{beam}\ll c$.
All Twiss parameters are evaluated at $z^{lab}=\gamma z^{beam} + \gamma \beta c t$.

We write the phase space density function (\ref{eq:f0x}) in terms of beam frame coordinates and velocities as:
\begin{equation}
f_x^0(x,v_x) = \frac{c_{\pi}}{\sigma_x} \exp\left[-\frac{(1+\alpha_x^2)x^2}{2r_x^2} - \frac{\alpha_x x v_x}{r_x\sigma_x}
- \frac{v_x^2}{2\sigma_x^2}\right]
\label{eq:f0x_beam}
\end{equation}
where $r_x$ and $\sigma_x$ are functions of $t$ and $z$,
because they are related to the Twiss parameters by the following formulas,
\begin{eqnarray}
r_x & = & \sqrt{\epsilon_x\beta_x} \label{eq:r_x} \\
\sigma_x & = & \gamma\beta c \sqrt{\frac{\epsilon_x}{\beta_x}}
\label{eq:sigma_x}
\end{eqnarray}
with the analogous formulas for the other transverse coordinate $y$.
We can think of $r_x$ and $\sigma_x$ as the current rms beam size and rms velocity.
Again, the (unsubscripted) $\beta$ and $\gamma$ in (\ref{eq:sigma_x}) are the relativistic invariants
coming from the conversion from lab frame to beam frame, and are unrelated to Twiss parameters.

In the longitudinal direction we have a Maxwellian distribution (in the beam frame),
\begin{equation}
f_z^0(z,v_z) = \frac{\density(z^{lab}) c_\pi}{\sigma_z} \exp\left[- \frac{v_z^2}{2\sigma_z^2}\right] ,
\end{equation}
where $v_z$ is the longitudinal particle velocity in the beam frame.
$\density(z^{lab})$ is the density at the center of the beam ($x=y=0$),
which is a function of time and $z$.
In terms of beam frame coordinates, $z^{lab}=\gamma z+ \gamma\beta c t$.

If we integrate the distribution $f^0$ over velocity space, we obtain the spatial distribution
\begin{eqnarray}
\eta^0(\v{x},t) & = & \int f^0(\v{x},\v{v},t) d\v{v} \nonumber \\
& =  & \density(\gamma z + \gamma\beta c t) \exp\left[-\frac{x^2}{2r_x^2} - \frac{y^2}{2r_y^2}\right] ,
\label{eq:fxy}
\end{eqnarray}
a density distribution which is Gaussian in $x$ and $y$
with time dependent rms widths $r_x$ and $r_y$.

To conserve charge in the beam cross section, the integral of $\eta^0$ over all $x$ and $y$ must be constant,
which implies that
\begin{equation}
\density(z^{lab}) = \density^0 \sqrt{\frac{\beta_x(z^{lab}=0) \beta_y(z^{lab}=0)}{\beta_x(z^{lab}) \beta_y(z^{lab})}}
\label{eq:n0}
\end{equation}
where, again $z^{lab}=\gamma z+ \gamma\beta c t$ and
$\density^0 = \density(z^{lab}=0)$ is the peak (central) density at the start of the simulation.
Thus, the $\density(z^{lab})$ is simply a function of the Twiss parameters.

\begin{figure}[htb]
\centering
\includegraphics[width=82mm]{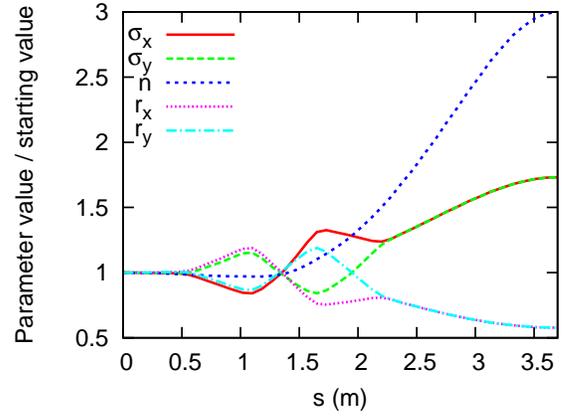}
\caption{Ratio of various electron beam parameters to their values at the start of the modulator.}
\label{fig:1.5}
\end{figure}

In Figure~\ref{fig:1} the horizontal and vertical beta functions go from 4.5 m/rad to 1.5 m/rad, corresponding to a decrease in
the beam radius by a factor of $\sqrt{3}$, and an increase in $\density$ by a factor of $3$ (Figure~\ref{fig:1.5}).
We use a peak density at the end of the modulator of $5.482\times 10^{16}$ $\text{e}/\text{m}^3$,
which corresponds to a starting density $\density^0 = 1.827\times 10^{16}$ $\text{e}/\text{m}^3$.

We use a normalized emittance of $3$ mm-mrad,
which corresponds to a rms emittance ${\epsilon_x=\epsilon_y=7.5\times10^{-8}}$~m-rad.
From (\ref{eq:r_x}) and (\ref{eq:sigma_x}), the starting values of the rms beam radius
${r_x=r_y=0.581}$~mm or 581~\micron~and temperature ${\sigma_x=\sigma_y=1.55\times10^6}$~m/s.
The longitudinal temperature is independent of Twiss parameters; we use a value of $\sigma_z=2.93\times 10^5$~m/s,
which corresponds to a lab frame momentum spread of ${\Delta p/p \approx 1\times 10^{-3}}$.
The Debye length $\lambda$ is a function of the temperature and density,
and therefore changes in time, as well as being smaller longitudinally than transversely.
Numerically, $\lambda_x = \sigma_x/\omega_p$, where $\omega_p$ is the local plasma frequency,
\begin{equation}
\omega_p = |e|\sqrt{\frac{n}{m\epsilon_0}}
\label{eq:plasmaf}
\end{equation}
At the start of the simulation, the transverse Debye length $\lambda_x$ is 203 \micron,
and the longitudinal Debye length $\lambda_z$ is 38 \micron.

By integrating equation (\ref{eq:fxy}) over all $x$ and $y$, we obtain
$2\pi\density r_x r_y$, the linear charge density in the beam frame, a constant in our model.
The lab-frame current associated with this linear charge density is 
\begin{equation}
I = 2\pi\density r_x r_y (\beta\gamma c e)
\end{equation}
Using the starting density $\density^0 = 1.827\times 10^{16}$ $\text{e}/\text{m}^3$ and the circular beam $r_x=r_y=0.581$ mm,
with the relativistic $\gamma=40.0$, we obtain a current of $I=74.4$ Amperes.
These parameters are summarized in Table~\ref{table0}.

\begin{table}[!t]
  \renewcommand{\arraystretch}{1.3}
  \centering
  \begin{tabular}{cccc} \hline
    & modulator & modulator & \\
    parameter & start & end & unit \\
    \hline
    peak density & $1.827$ & $5.482$ & $10^{16} \text{e}/\text{m}^3$ \\
    current & $74.4$ & $74.4$ & Amperes \\
    emittance & 3.0 & 3.0 & mm-mrad \\
    plasma freq. & 1.214 & 2.103 & $10^9$/sec \\
    plasma per. & 0.824 & 0.476 & nanoseconds \\
    $r_x$, $r_y$ & $581$ & $335$ & \micron \\
    $\sigma_x$, $\sigma_y$ & $1.55$ & $2.68$ & $10^6$ m/sec \\
    $\sigma_z$ & $0.29$ & $0.29$ & $10^6$ m/sec \\
    $\lambda_x, \lambda_y$ & $203$ & $203$ & \micron \\
    $\lambda_z$ & $38$ & $22$ & \micron \\
    \hline
  \end{tabular}
  \caption{\label{table0}Electron beam parameters in the modulator.
All parameters are in the beam frame except for the current.}
\end{table}

In order to specify the weight equations (\ref{eq:wt2}) and (\ref{eq:wt3}),
we calculate $\grad_p f^0 = (1/m)\grad_v f^0$ directly from the distribution function (\ref{eq:f0})
and (\ref{eq:f0x_beam}),
\begin{equation}
\frac{\grad_p f^0}{f^0} = -\frac{1}{m} \left(\frac{\alpha_x x}{r_x\sigma_x}+\frac{v_x}{\sigma_x^2},
\frac{\alpha_y y}{r_y\sigma_y}+\frac{v_y}{\sigma_y^2},\frac{v_z}{\sigma_z^2}\right) 
\label{eq:f0v_beam}
\end{equation}
The expression (\ref{eq:f0v_beam}) is very similar to the case where $f^0$ is uniform in space,
adding only the terms involving $\alpha_x$ and $\alpha_y$.

\section{An exact shielding solution}
Wang and Blaskiewicz \cite{WB08} discovered an exact solution for shielding,
assuming the electron density $f^0$ is uniform in space and extends to infinity,
and has a special ``$\kappa$-2'' velocity distribution.
Here we work in the ion reference frame where the ion is stationary.
This differs from the beam-frame only by the addition $\v{v^i}$ of the ion velocity.
Since the ion speed in the beam frame is non-relativistic,
we can move between the beam and ion frames using a Galilean transformation.
Any changes to the ion velocity due to the background and shielding fields are
small---we can assume the ion drifts with constant velocity.

In the ion frame the special ``$\kappa$-2" electron velocity distribution is given by
\begin{eqnarray}
f^0(\v{v}) & = & \frac{n_0}{\pi^2\wbbeta_x\wbbeta_y\wbbeta_z}\left[1+\left|\frac{\v{v}+\v{v^i}}{\v{\wbbeta}}\right|^2\right]^{-\kappa}
\label{eq:kappa2} \\
 & = & \frac{n_0}{\pi^2\wbbeta_x\wbbeta_y\wbbeta_z}\left[1+\frac{(v_x+v^i_x)^2}{\wbbeta_x^2}+\frac{(v_y+v^i_y)^2}{\wbbeta_y^2}\right. \nonumber \\
 &  & + \left.\frac{(v_z+v^i_z)^2}{\wbbeta_z^2} \right]^{-\kappa} \nonumber
\end{eqnarray}
here $\kappa=2$ and
$\v{\wbbeta}=(\wbbeta_x,\wbbeta_y,\wbbeta_z)$ define the width of the velocity distribution in $x$, $y$ and $z$,
they are analogous to the rms values $\v{\sigma}$ in the Gaussian case.
Wang and Blaskiewicz \cite{WB08} use $\beta$ in (\ref{eq:kappa2}) in place of $\wbbeta$,
however we are already using $\beta$ as the relativistic constant,
as well as $\beta_x$ and $\beta_y$ for Twiss parameters.
In (\ref{eq:kappa2}), the division of $\v{v}+\v{v^i}$ by $\v{\wbbeta}$ is to be performed component-wise.

Note that the 3D $\kappa$-2 distribution (\ref{eq:kappa2}) cannot be separated into the product of three functions,
one for each dimension, as is the case for the Gaussian (\ref{eq:f0}).
If we integrate (\ref{eq:kappa2}) over all $v_z$ we get the 2D form
\begin{equation}
f^0(\v{v}) = \frac{n_0}{2\pi\wbbeta_x\wbbeta_y}\left[1+\left|\frac{\v{v}+\v{v^i}}{\v{\wbbeta}}\right|^2\right]^{-3/2}
\label{eq:kappa2D}
\end{equation}
and integrating this over all $v_y$ gives the 1D form
\begin{equation}
f^0(v_x) = \frac{n_0}{\pi\wbbeta_x}\left[1+\left|\frac{v_x+v_x^i}{\wbbeta_x}\right|^2\right]^{-1}
\label{eq:kappa1D}
\end{equation}

\begin{figure}[htb]
\centering
\includegraphics[width=82mm]{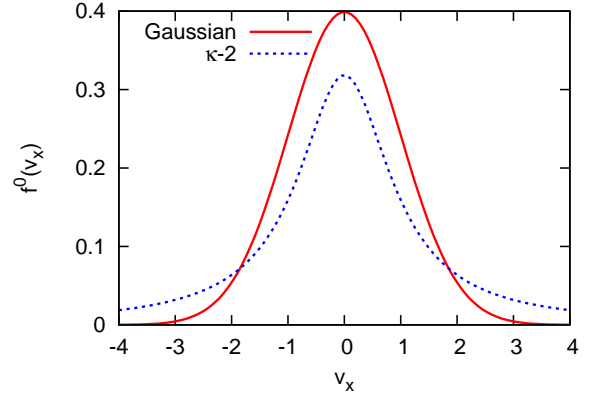}
\caption{A comparison between the Gaussian and $\kappa$-2 distribution functions in 1D.}
\label{fig:2}
\end{figure}

Figure~\ref{fig:2} shows a 1D comparison between the Gaussian and $\kappa$-2 distribution (\ref{eq:kappa1D}),
with the widths set to 1 ($\sigma_x=\wbbeta_x=1$).
Note that the 1D $\kappa$-2 distribution (\ref{eq:kappa1D}) decays as $v_x^{-2}$ for large $v_x$,
while the Gaussian distribution decays exponentially.

Let $\eta^1$ be the exact solution for the perturbation $f^1$ integrated over velocity space, i.e.
\begin{equation}
\eta^1(\v{x},t) = \int f^1(\v{x},\v{v},t) d\v{v}
\end{equation}
then the exact solution for $\eta^1$ derived by Wang and Blaskiewicz \cite{WB08} is given by the single integral
\begin{equation}
\eta^1(\v{x},t) = \frac{Z}{\pi^2}\int\limits_0^t
\frac{\omega_p s\sin(\omega_p s) ds}{\wbbeta_x\wbbeta_y\wbbeta_z\left[s^2 + \left|(\v{x}+s\v{v^i})/\v{\wbbeta}\right|^2\right]^2}
\label{eq:exactsol}
\end{equation}
Again, the division of two vectors in (\ref{eq:exactsol}) is to be performed component-wise.

To compare with simulations, we will plot not the 3D distribution $\eta^1(\v{x},t)$ but integrate it over two dimensions
and plot either as a line curve or a contour plot over time.
For example, if we integrate over $y$ and $z$ we can define
\begin{equation}
\eta^1(x,t) = \int_{-\infty}^{\infty} \int_{-\infty}^{\infty} \eta^1(\v{x},t)\,dy\:dz
\label{eq:eta1xdef}
\end{equation}
Now if we insert the formula (\ref{eq:exactsol}), we can switch the order of integration and
calculate the integrals over $y$ and $z$, leaving a single integral over time,
\begin{equation}
\eta^1(x,t) = \frac{Z}{\pi}\int\limits_0^t
\frac{\omega_p s\sin(\omega_p s) ds}{\wbbeta_x(s^2 + \left|(x+s v_x^i)/\wbbeta_x\right|^2)} .
\label{eq:exactsol2}
\end{equation}
The integration in (\ref{eq:eta1xdef}) can also be done over the finite simulation domain,
although the formula (\ref{eq:exactsol2}) becomes more complicated.

In the exact solution (\ref{eq:exactsol}), the plasma frequency $\omega_p$ and $\v{\wbbeta}$ are constants.
In our more complex beam defined by Twiss parameters, the plasma frequency $\omega_p$,
as well as the electron temperature parameter $\v{\wbbeta}$ vary in space and time.
The integrand in (\ref{eq:exactsol}) can be interpreted as the contribution to the shielding wake at time $s$.
Hence it makes sense in this integrand to substitute the time varying values of $\omega_p$ and $\v{\wbbeta}$
at the location of the ion.
After the integral is calculated (numerically), we get an estimate for the wake created by an ion
inside a beam where the local density is changing.

In our simulations we also use a Gaussian velocity distribution rather than a $\kappa$-2 distribution,
but (\ref{eq:exactsol}) gives us an estimated solution by substituting the rms value $\v{\sigma}$ for $\v{\wbbeta}$. 
Of course $\sigma_x$ and $\sigma_y$ are functions of the current Twiss parameters as defined by (\ref{eq:sigma_x}).
One of the few parameters that does not vary in our beam is the longitudinal parameter $\sigma_z$.
Given all the differences between the exact model and the realistic beam model,
we expect results to match only qualitatively.
The advantage of the exact formula (\ref{eq:exactsol}) is that is easy to calculate.

Although the exact solution (\ref{eq:exactsol}) is for a simplified model,
it shares several important features with our realistic beam.
The electron distribution can be anisotropic, and the ion can be moving in an arbitrary direction.
In Appendix~A it is shown that for the exact solution, the total shielding charge $Q$ as
a function of time is $Q = Z(1 - \cos \omega_p t)$, see equation (\ref{eq:scharge}).

\section{Simulation Results}

\begin{figure}[htb]
\centering
\includegraphics[width=82mm]{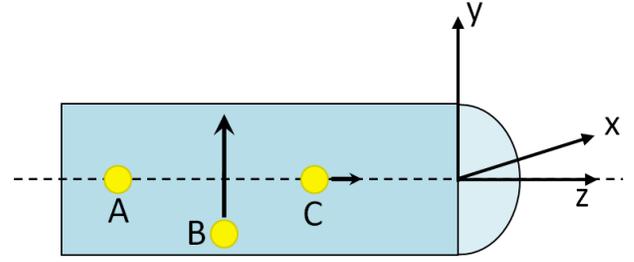}
\caption{Electron beam schematic, showing the three simulated ions. $z$ is the direction of beam propagation.}
\label{fig:diagram}
\end{figure}

We simulate shielding of three specific gold ions through the modulator (see Figure~\ref{fig:diagram}):
\begin{enumerate}[A.]
\item An ion which is stationary at the center of the beam.
\item An ion which moves transversely at speed $2.68\times 10^6$ m/s (the electron transverse thermal speed).
The initial position is chosen so that the ion passes the beam center when half way through the modulator.
This ion moves from one side of the beam to the opposite side during the simulation.
\item An ion which moves longitudinally at speed $2.93\times 10^5$ m/s (the electron longitudinal thermal speed).
\end{enumerate}

\begin{figure}[htb]
\centering
\includegraphics[width=82mm]{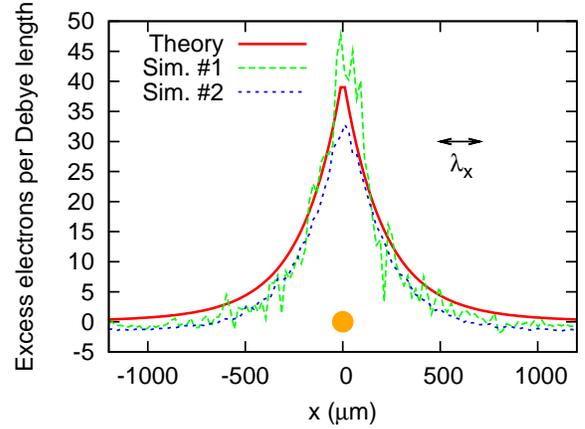}
\caption{Transverse shielding of ion `A' at the end of the modulator compared with theory.
The gold dot marks the ion.}
\label{fig:10}
\end{figure}

In order to compare with the exact results of Wang and Blaskiewicz \cite{WB08},
we start with a simplified beam model.
Specifically, we assume the Twiss parameters are fixed at their values at the exit to the modulator.
This results in a beam which does not change in time.
The density function (\ref{eq:f0}) is constant in time, but varies in space.
In order that the beam be in equilibrium, we would need some kind of external focusing field.

In particular, we will use four types of simulations:
\begin{enumerate}
\item Twiss parameters fixed, $\delta$f particles distributed uniformly in phase space.
Particle weights evolve according to Equation (\ref{eq:wt2}).
\item Twiss parameters fixed, $\delta$f particles distributed as in the actual beam.
Particle weights evolve according to Equation (\ref{eq:wt3}).
\item Twiss parameters as in the real beam, $\delta$f particles distributed uniformly in phase space.
\item Twiss parameters as in the real beam, $\delta$f particles distributed as in the actual beam.
\end{enumerate}
In what follows we will refer to a simulation in the above list as `Sim. \#2', for example.

\begin{figure}[htb]
\centering
\includegraphics[width=82mm]{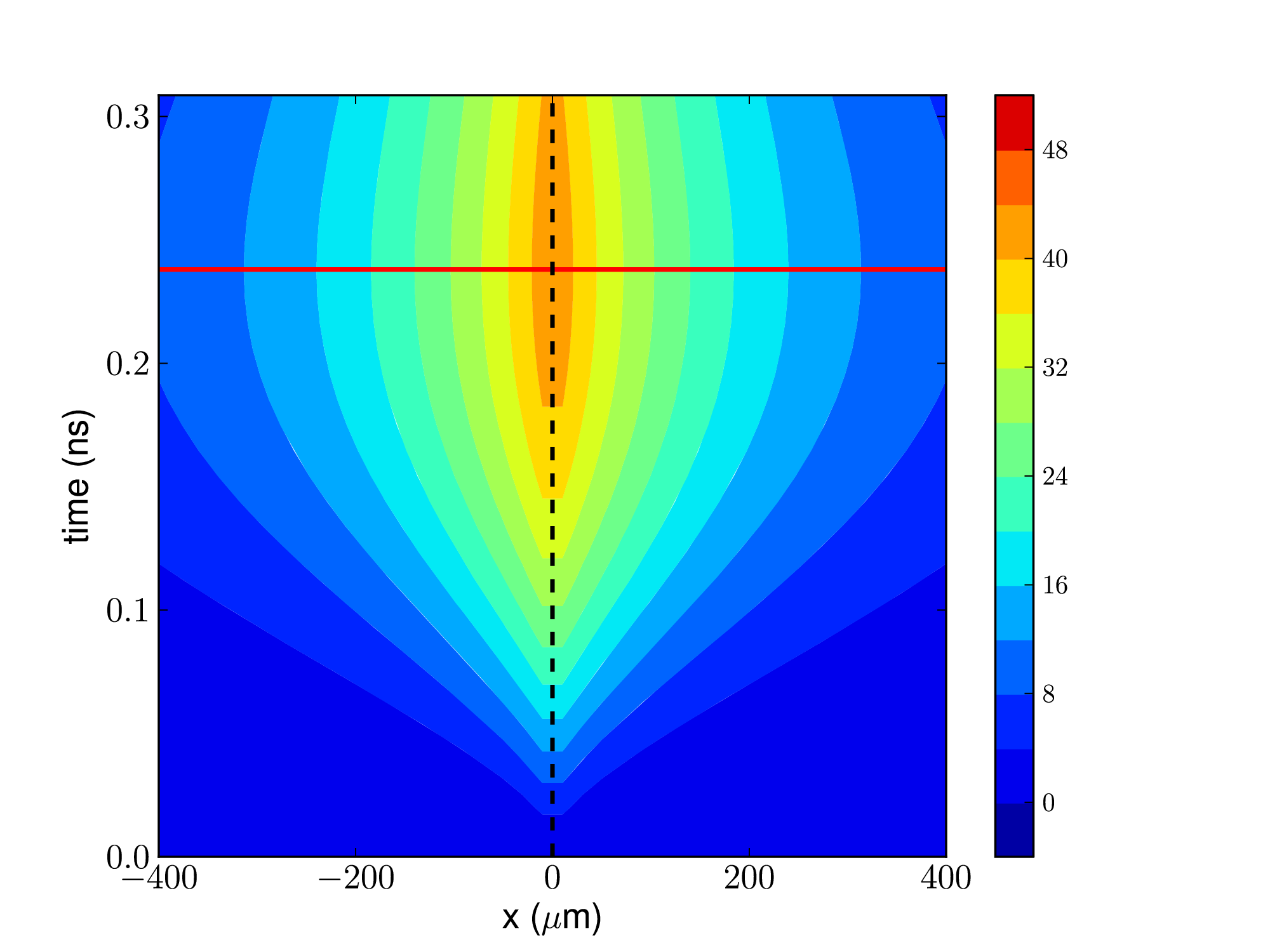}
\caption{Ion `A', theoretical shielding.  The horizontal red line marks half a plasma period, where the response is maximum.
The vertical dashed line marks the ion trajectory.}
\label{fig:11b}
\end{figure}

\begin{figure}[htb]
\centering
\includegraphics[width=82mm]{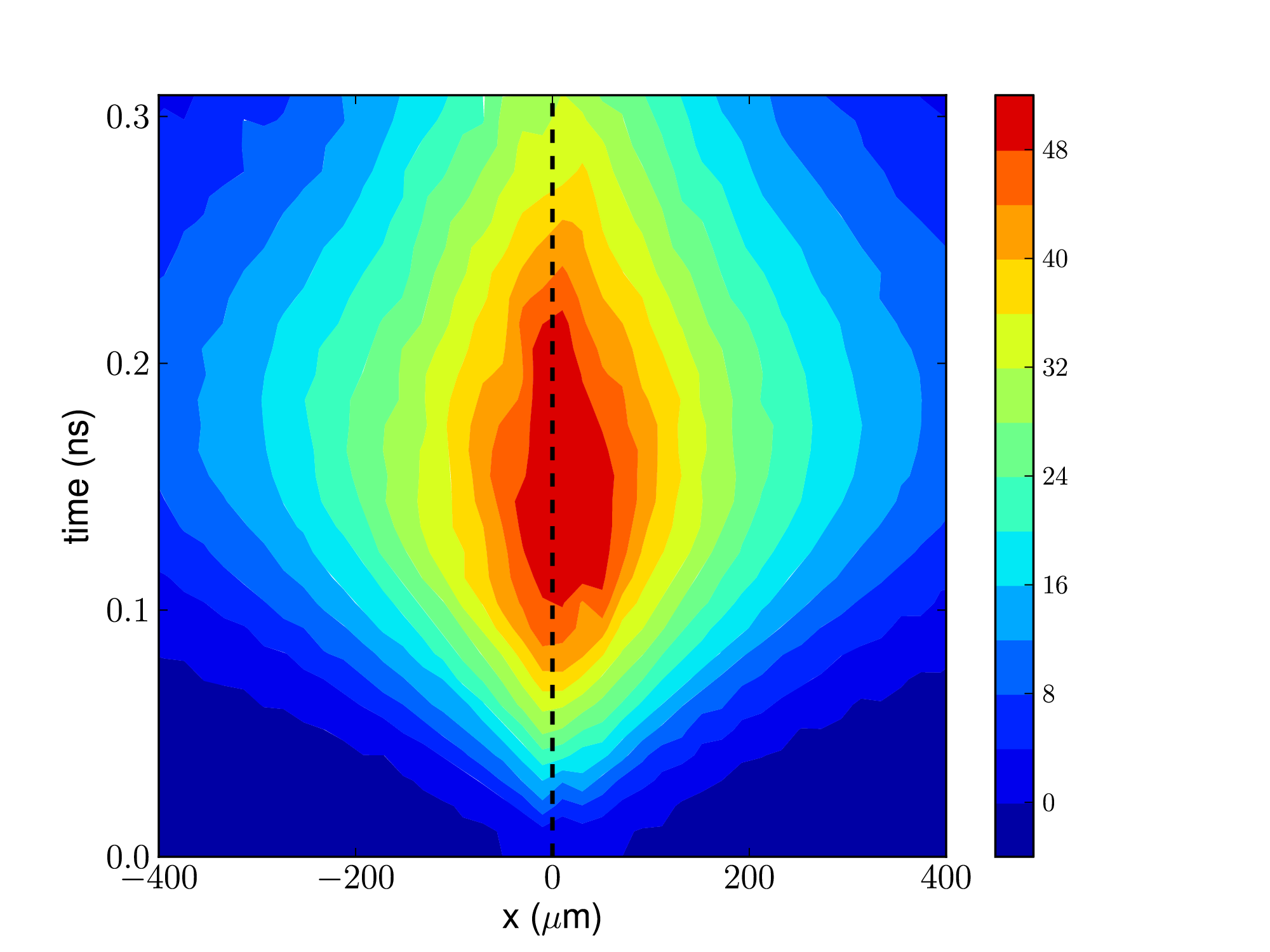}
\caption{Ion `A', Sim. \#2.}
\label{fig:11c}
\end{figure}

\begin{figure}[htb]
\centering
\includegraphics[width=82mm]{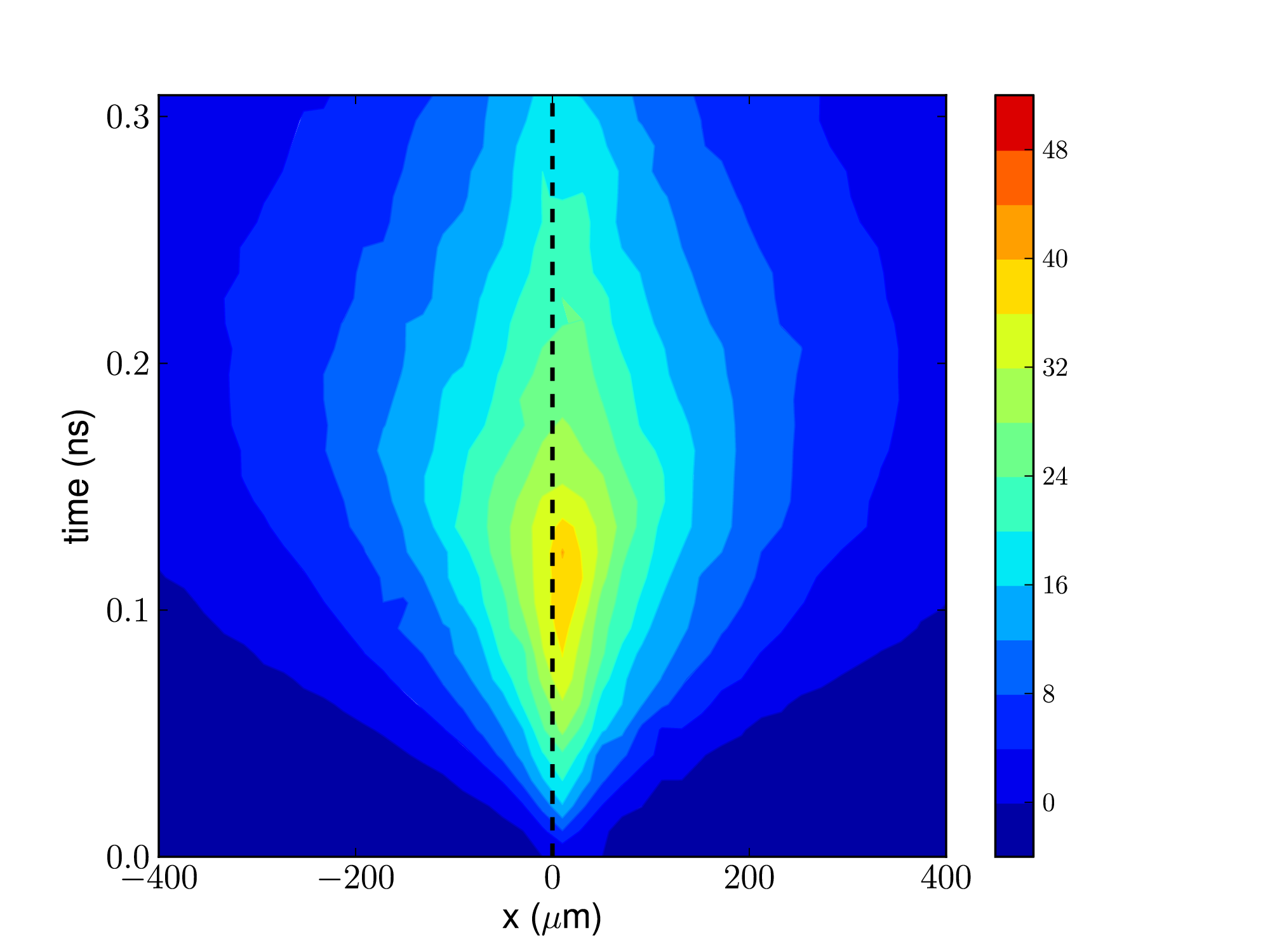}
\caption{Ion `A', Sim. \#4.}
\label{fig:11d}
\end{figure}

For Figures~\ref{fig:10} to \ref{fig:11d},
we place a stationary gold ion in the center of the beam, where the density is approximately constant.
We use a Gaussian velocity distribution rather than the $\kappa$-2 distribution specified by the exact solution.
The theoretical solution assumes an electron distribution which is uniform in space and infinite in extent.
Nonetheless, we can see that the agreement between the Wang and Blaskiewicz \cite{WB08} solution and 
the numerical solutions is qualitatively quite good (Figure~\ref{fig:10}).
The color scale ``excess electrons" is the density $\eta(x,t)$
which has been integrated in the other two dimensions (\ref{eq:eta1xdef}).

One of the subtlest aspects of the numerical simulations are the boundary conditions.
We have boundary conditions for the Poisson solve, and boundary conditions for the $\delta$f particles.
Our computational box will have periodic boundary conditions in $z$ (the longitudinal dimension),
because we are assuming at the Debye length scale the beam is uniform in $z$.
Transversely the density goes to zero exponentially, so our computational domain need only be
on the order of $6r_x$ wide.
For the Poisson solve, we use periodic boundary conditions in $z$, and Dirichlet boundary conditions in $x$ and $y$.
We also have tried using special open boundary conditions in $x$ and $y$, but this does not seem to change the results.
The $\delta$f particles have periodic boundary conditions in $z$, but are lost if they hit the domain boundary in $x$ or $y$.
The weight of these particles is lost, so they total weight over all particles is not conserved in our simulations.

Figures~\ref{fig:11b} to~\ref{fig:11d} show the shielding response as a color contour map,
where the vertical axis is time.  The ion trajectory (dashed line) is vertical because the
ion is stationary.
The horizontal axis in these plots is $x$, and the density has been integrated over the other two coordinates.
For these integrations, we have not used the entire computational domain, but only a subset of the domain,
to remove any boundary effects.

Because the longitudinal Debye length $\lambda_z$ is nearly 10 times smaller than the transverse Debye length $\lambda_x$,
our grid spacing in $z$ must be much smaller than in $x$ and $y$.
To resolve the isotropic and slowly decaying field of the ion, we need to include many more grid points in $z$
than in $x$ and $y$.
A typical grid for these simulations is $140\times 140\times 500$.

The focusing quadrupoles present a problem for our simulations \#3 and \#4 of the full Twiss beam.
These quadrupoles focus the bulk Twiss beam, and are already included in the base solution $f^0$.
However, these quadrupoles should also act on the $\delta$f particles, focusing them as well.
The transverse velocity change of an electron passing $1$ mm off center through the
quadrupole can be estimated using the electric field in the beam frame (\ref{eq:ebeam}),
and the $4$ mm beam frame quadrupole length.
The result is a transverse velocity kick on the order of $c/4$.
These large velocity kicks violate our assumption that particle velocities are nonrelativistic in the beam frame.
They also present significant problems for the numerical integration of particle trajectories,
as very small time steps will be needed to resolve the large accelerations.
We note that in the lab frame, transverse electron velocities are reduced by a factor of $\gamma=40$,
so they are not relativistic!

\begin{figure}[htb]
\centering
\includegraphics[width=82mm]{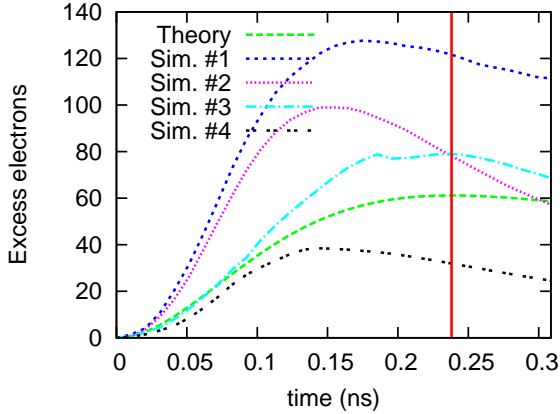}
\caption{Shielding charge within 2 Debye lengths of the ion `A' as a function of time.
The vertical red line marks half a plasma period for the central (maximum) density.}
\label{fig:12zz}
\end{figure}

Our simulation results therefore do not include the effect of the quadrupoles on the $\delta$f particles,
which should result in some reduction of the ion shielding (due to the lack of focusing).
Figure~\ref{fig:11d} shows that the shielding response is only about half the amount of
that in Simulations \#1 and \#2, where the Twiss parameters are fixed.

Figure~\ref{fig:12zz} shows the total shielding charge within 2 Debye lengths of the stationary ion.
By (\ref{eq:scharge}) the theoretical shielding should go as $Z(1-\cos\omega_p t)$, which gives $2Z=158$ at half a plasma period.
In Figure~\ref{fig:12zz} the theoretical peak is more like $60$, the reason for the difference is that
it is only the charge within 2 Debye lengths, to obtain $2Z$ one would have to go out to infinity.
One can see from Figure~\ref{fig:13} (in Appendix~A) that less than half of the shielding
charge at $1/2$ plasma period is within 2 Debye lengths of the ion.

The simulations in Figure~\ref{fig:12zz} show a curious effect, namely that the maximum shielding response
appears \textit{before} half a plasma period.
This seems counterintuitive, because that half plasma period is based on the maximum density of the beam.
Most of the time, and away from the center of the beam, the electron density is lower, and therefore the local
plasma period is longer, so one would expect, if anything, that the peak shielding would occur later than half a plasma period.
Note, however, that we have introduced a faster time scale, which is the time scale over which the
Twiss parameters, and therefore the beam, are varying (Figures~\ref{fig:1} and~\ref{fig:2}).

\begin{figure}[htb]
\centering
\includegraphics[width=82mm]{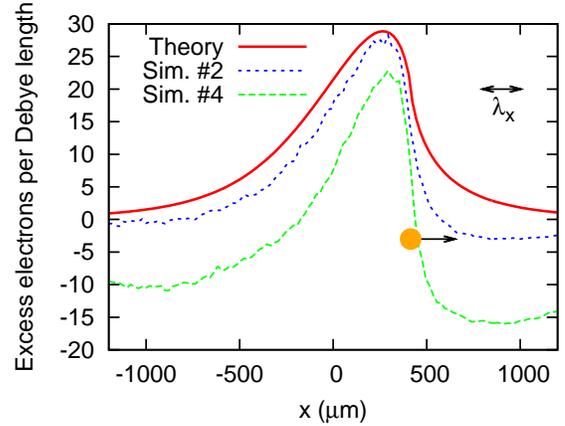}
\caption{Transverse shielding of ion `B' compared with theory.
The gold dot marks the moving ion.}
\label{fig:11}
\end{figure}

\begin{figure}[htb]
\centering
\includegraphics[width=82mm]{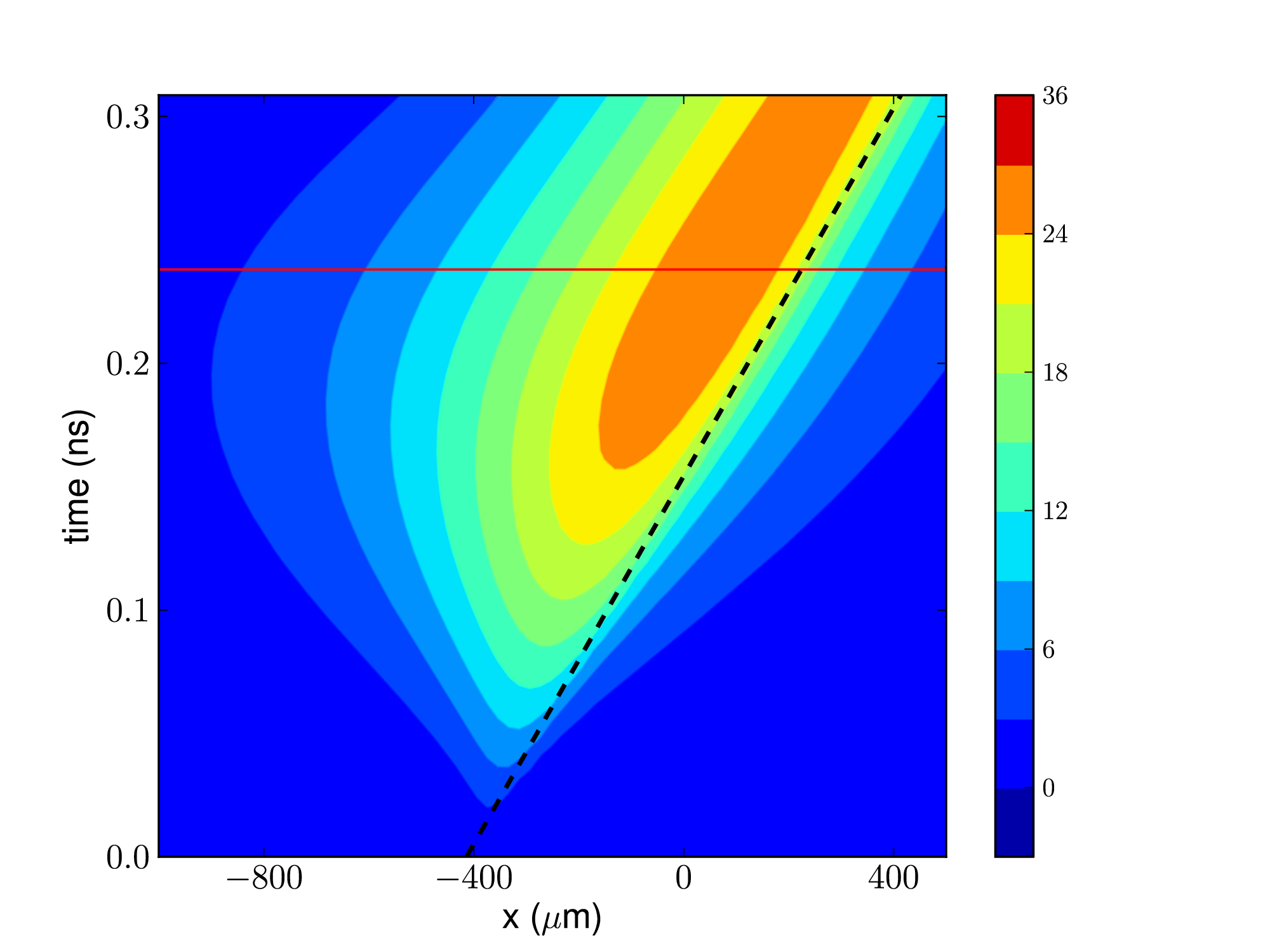}
\caption{Contour plot of the theoretical shielding for Ion `B' as a function of time.
The horizontal red line marks half a plasma period, where the response is maximum. The black dashed line marks the ion trajectory.}
\label{fig:12y}
\end{figure}

\begin{figure}[htb]
\centering
\includegraphics[width=82mm]{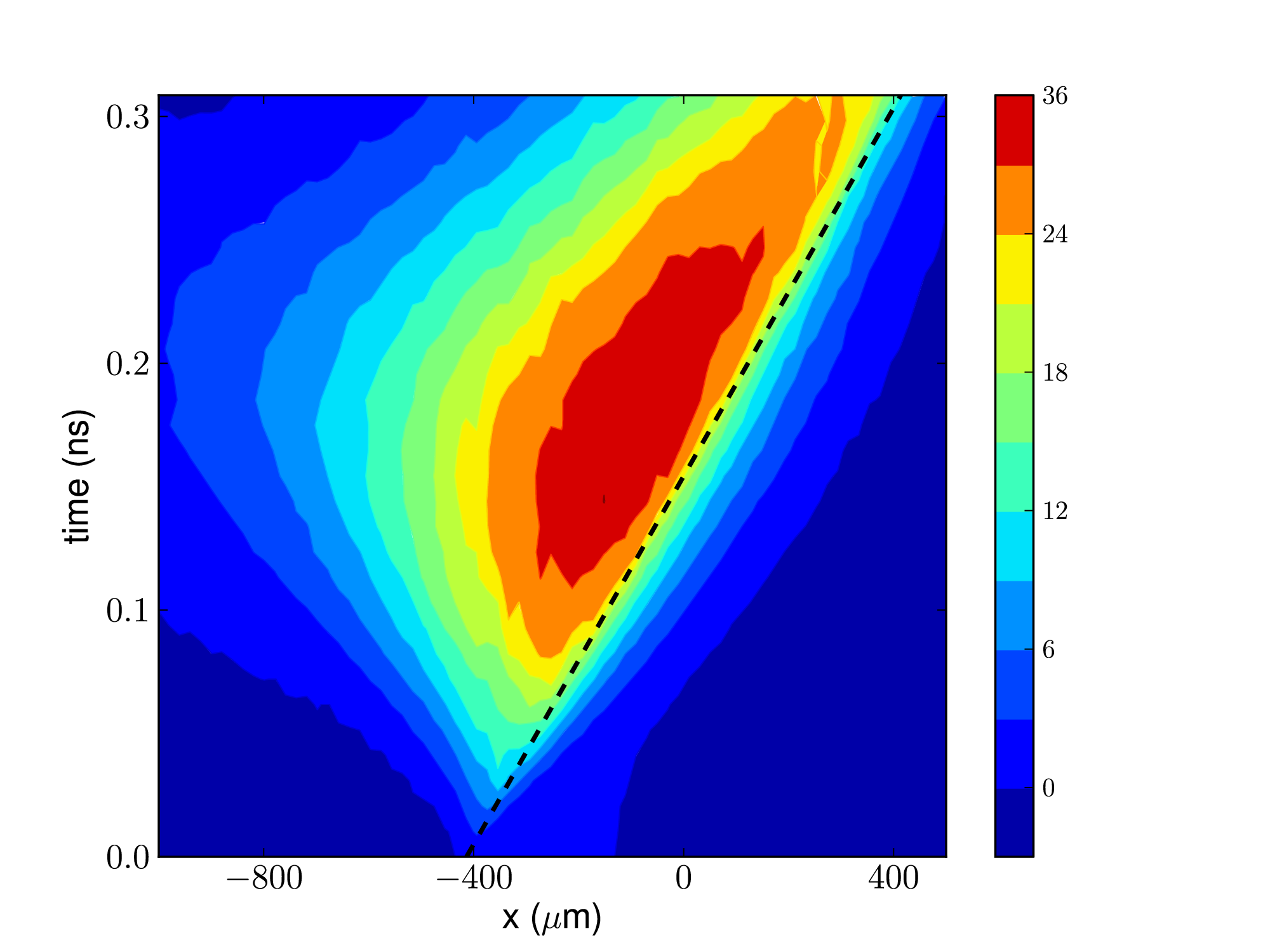}
\caption{Contour plot of the Sim. \#2 shielding for Ion `B' as a function of time.}
\label{fig:12z}
\end{figure}

\begin{figure}[htb]
\centering
\includegraphics[width=82mm]{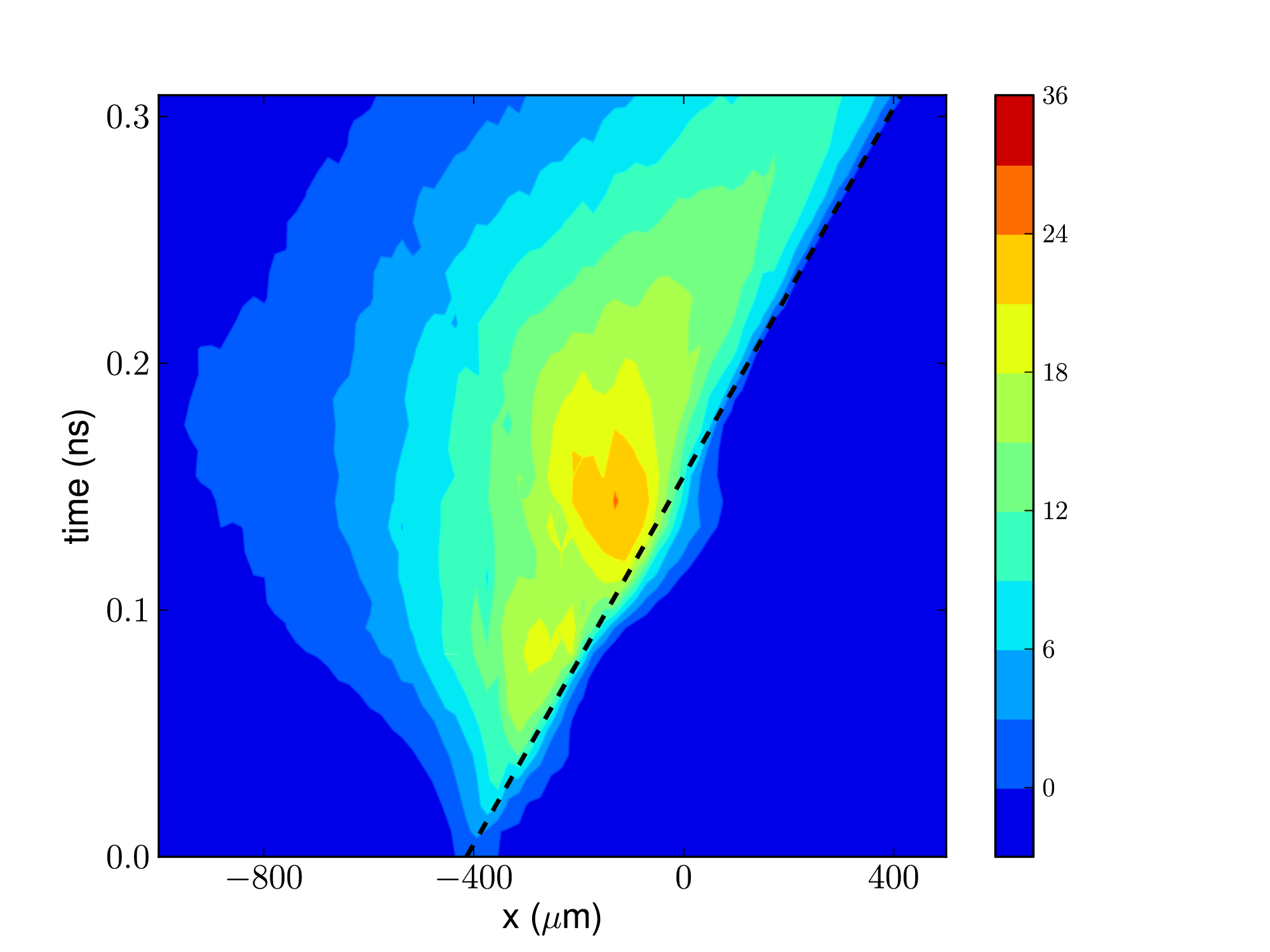}
\caption{Contour plot of the Sim. \#4 shielding for Ion `B' as a function of time.}
\label{fig:12zzz}
\end{figure}

In Figure~\ref{fig:11} we show the shielding response to a moving ion.
Ion `B' is moving transversely at the thermal speed $2.68\times10^6$ m/s, because of its transverse motion
the ion sees a varying electron density even though the Twiss parameters are constant.
At the start of the modulator, the ion has $x$ position $-414$ \micron, and by the end of the modulator
(Figure~\ref{fig:11}) its $x$ position is $+414$ \micron.

Figures~\ref{fig:12y} to~\ref{fig:12zzz} show color contour plots of the response
of moving ion `B'.  The dashed lines show the ion trajectory.
Again the peak shielding occurs before half a plasma period for the simulations,
the response for Sim. \#2 is stronger than the theoretical (constant density) shielding, while
Sim. \#4 is weaker and more similar to the theoretical shielding, although it peaks earlier.
It is easier to understand the early peaking of the response in this case, because
the density near the ion is highest in the middle of the simulation, when it passes through the
center of the beam.  Thus, one might expect the strongest response half-way through the
modulator, which is what is seen in Figures~\ref{fig:12z} and ~\ref{fig:12zzz}.

\begin{figure}[htb]
\centering
\includegraphics[width=82mm]{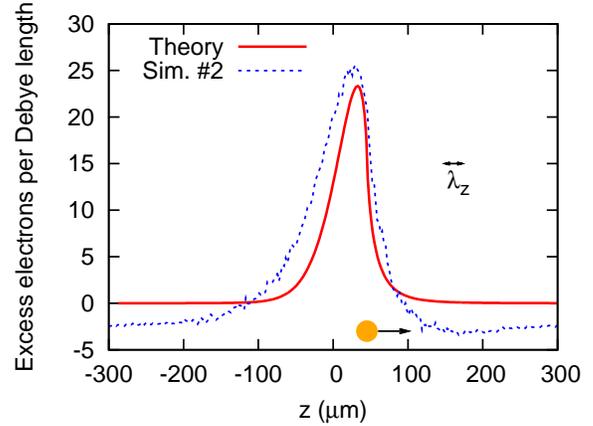}
\caption{Longitudinal shielding of ion `C' at the end of the modulator compared with theory.
The gold dot marks the moving ion.}
\label{fig:14a}
\end{figure}

\begin{figure}[htb]
\centering
\includegraphics[width=82mm]{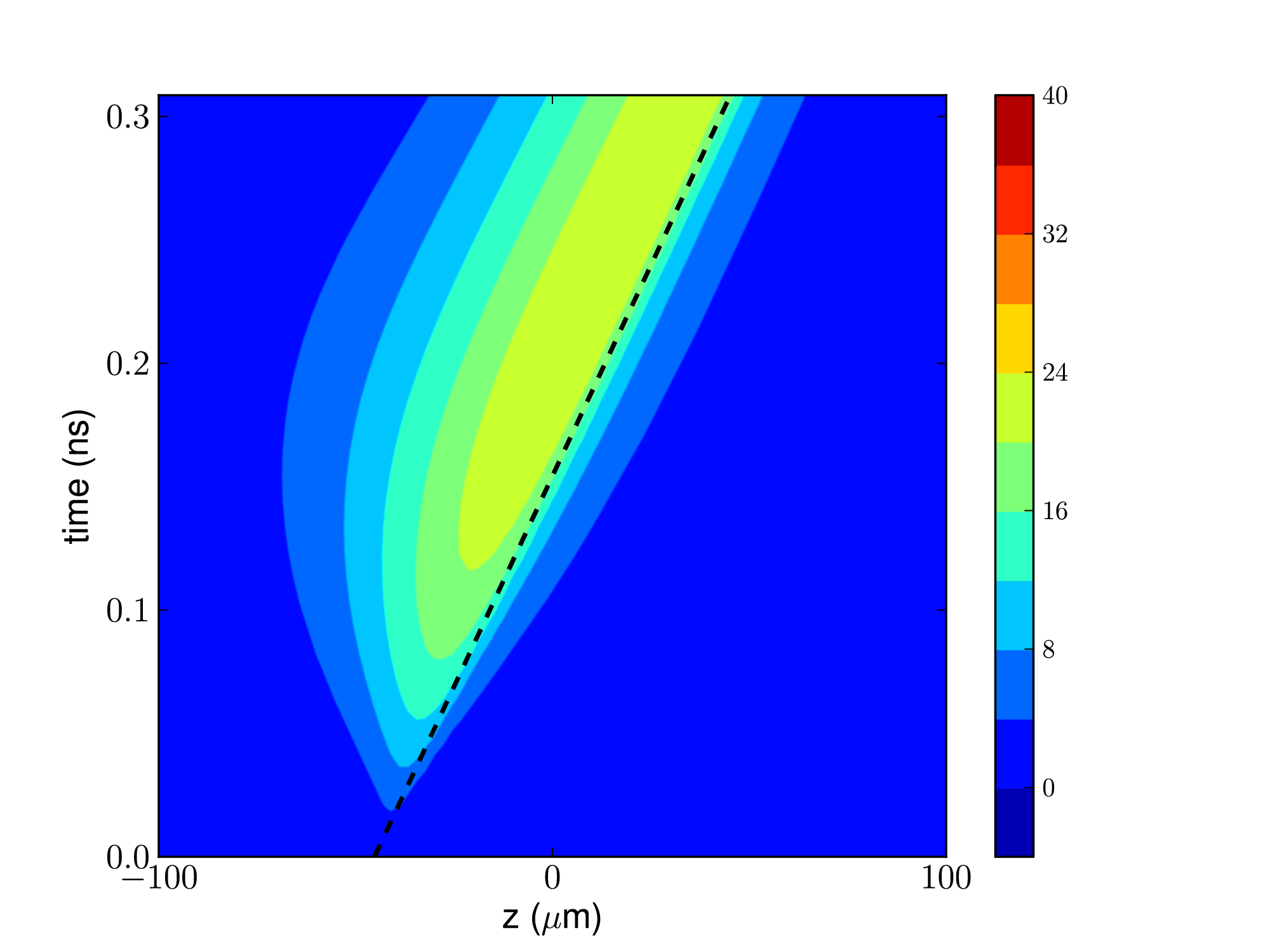}
\caption{Longitudinal shielding of ion `C' according to theory.}
\label{fig:14b}
\end{figure}

\begin{figure}[htb]
\centering
\includegraphics[width=82mm]{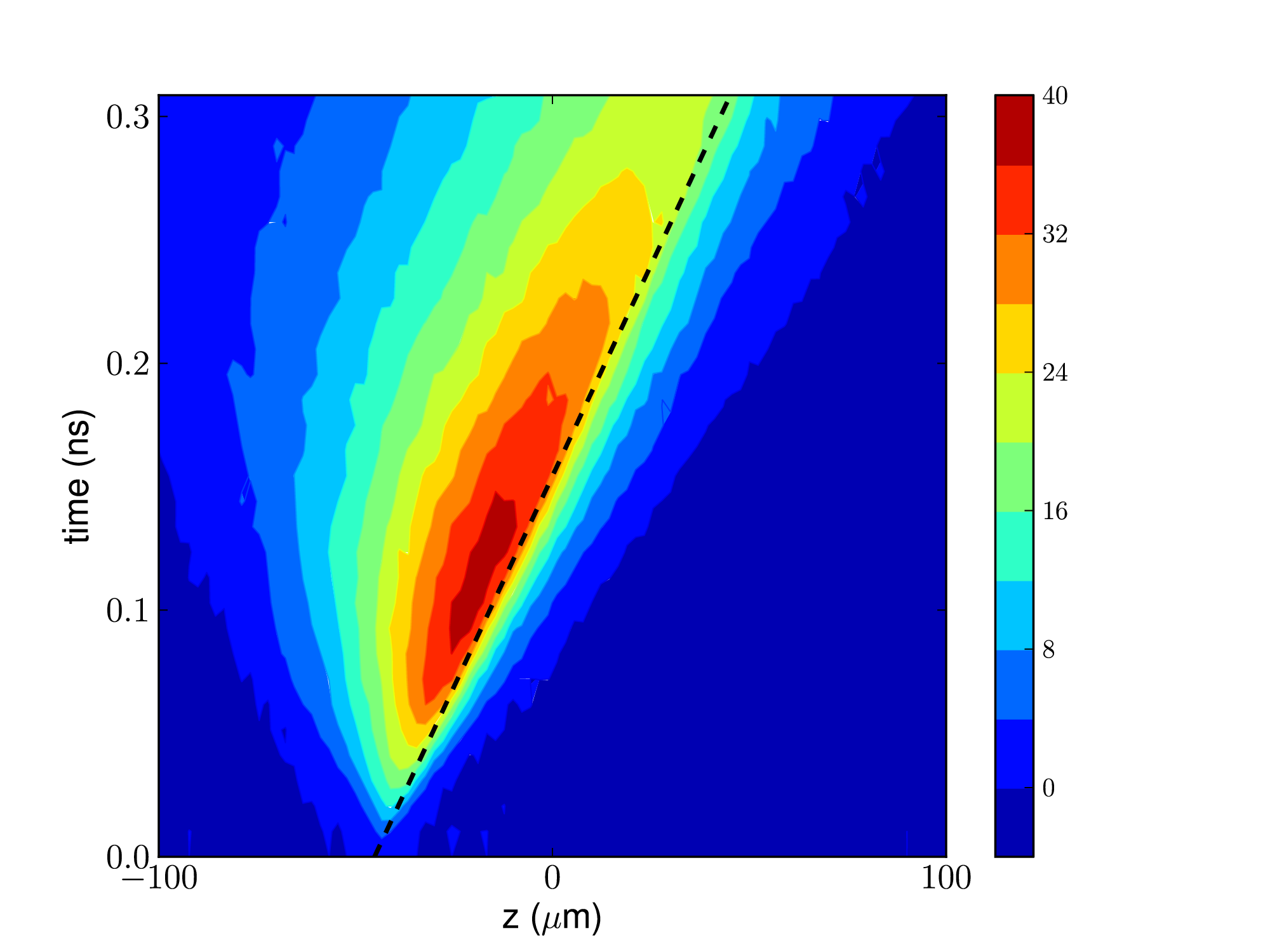}
\caption{Longitudinal shielding of ion `C', Sim. \#2.}
\label{fig:14c}
\end{figure}

\begin{figure}[htb]
\centering
\includegraphics[width=82mm]{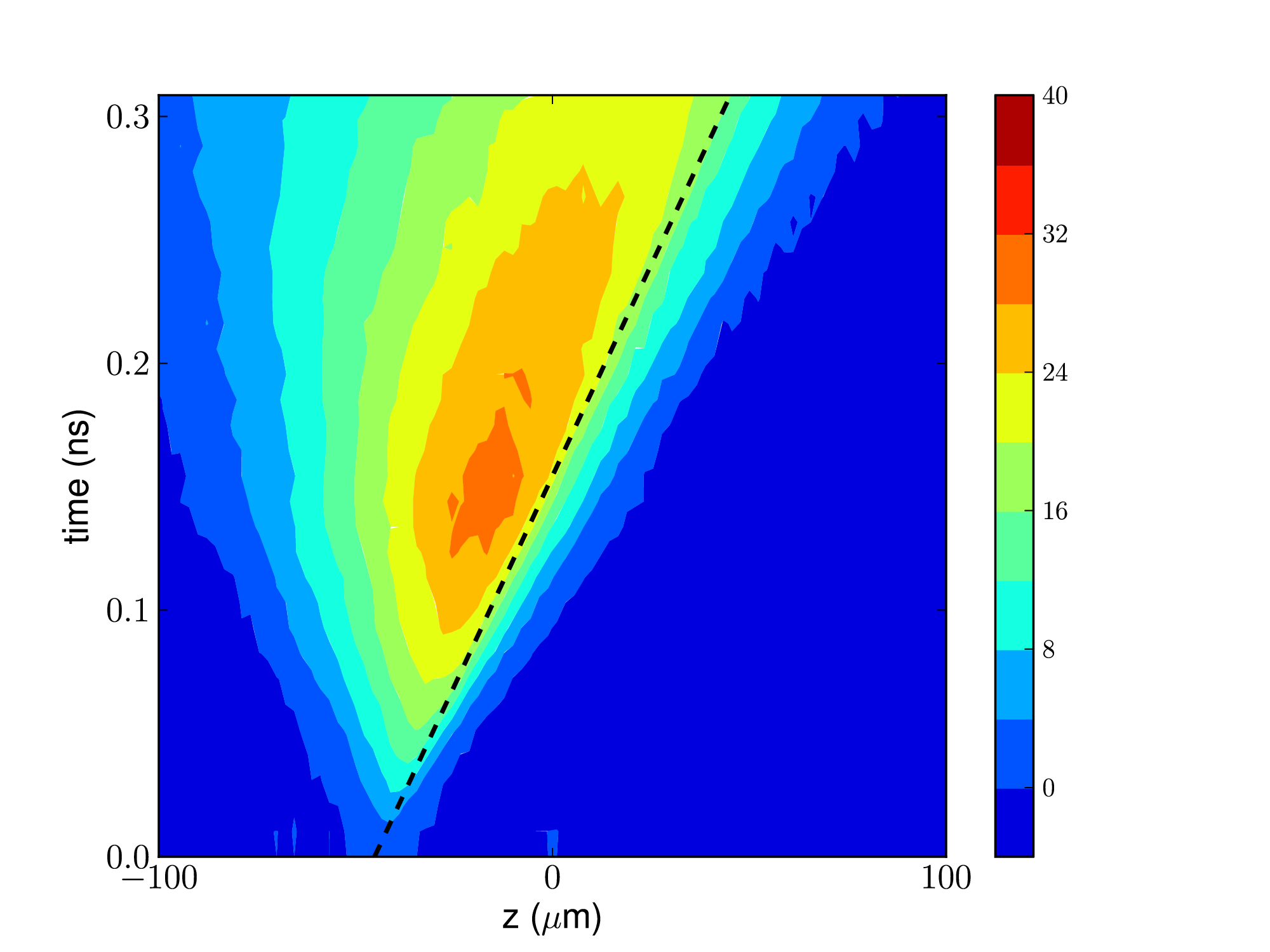}
\caption{Longitudinal shielding of ion `C', Sim. \#4.}
\label{fig:14d}
\end{figure}

Figure~\ref{fig:14a} shows the longitudinal shielding of an ion which is moving longitudinally along the
center of the beam.
The density seen by the ion is constant, so the plasma frequency is constant in this simulation.
The ion is moving at the longitudinal thermal speed, $2.9\times10^5$ m/s.
The shielding predicted by VSim is somewhat stronger than that given by the theoretical model,
but the agreement is still quite good.

Color contour plots of the density are shown in Figures~\ref{fig:14b} to~\ref{fig:14d}.
Again we see a stronger response in Sim. \#2 which peaks earlier, but a weaker response in
Sim. \#4 which is closer to the theoretical (constant density) case.
The horizontal scale is much smaller for these plots because the longitudinal Debye length is much
smaller than the transverse Debye length.



\section{Summary}

We have presented simulations of ion shielding which account for an electron beam which is focused in time,
and where the electron density goes to zero transversely.
The bulk of the beam is described by the Twiss parameters, while the ion shielding perturbation
has been represented by $\delta$f particles.
The maximum shielding charge often appears at a time somewhat less than half a plasma period,
calculated based on the peak density in the beam.

The constant-density theory of Wang and Blaskiewicz \cite{WB08} qualitatively produces similar results.
Nonetheless, the simulations show differences in the timing and magnitude of the shielding.
One would expect that Simulations \#1 and \#2 would match the theory more closely,
because for them at least the maximum electron density in the middle of the beam is constant in time.
However, the simulations show a stronger shielding response compared to theory,
which peaks earlier than half a plasma period.
Simulations \#3 and \#4 match the constant density theory better in the magnitude of the shielding,
although it also seems to peak earlier.

In simulating the entire Coherent Electron Cooling process, the next step
is to take the electron density perturbations at the end of the modulator,
and run them through an FEL simulation,
and finally the kicker simulation where the amplified perturbation interacts with the ion.
These results are discussed in other papers \cite{Schwartz13}.

\section{Acknowledgement}
The authors would like to thank the VSim development team and 
the BNL Collider Accelerator Division, especially Igor Pinayev.

\section{Appendix A: Calculation of the shielding charge}
\begin{figure}[htb]
\centering
\includegraphics[width=82mm]{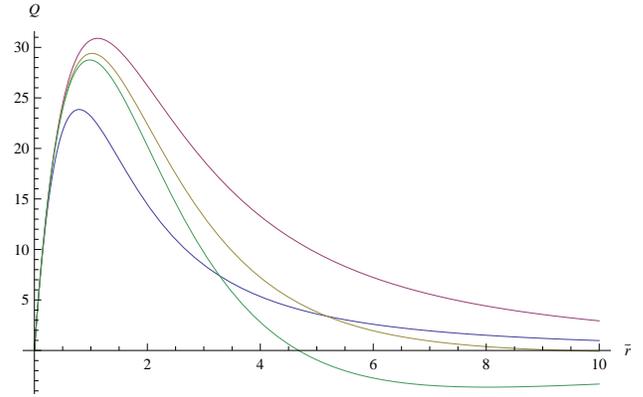}
\caption{Shielding charge as a function of $\bar{r}$ for a gold ion with $Z=79$ for
1/4, 1/2, 3/4 and 1 plasma periods (blue, red, brown, green).}
\label{fig:13}
\end{figure}

For the constant-density theory of Wang and Blaskiewicz, we can calculate the amount of shielding charge
within a certain distance of the ion, as well as the total shielding charge as a function of time.
To calculate the shielding charge at time $t$ for a stationary ion ($\v{v^i}=\v{0}$),
we integrate (\ref{eq:exactsol}) over all space.
To do this we define nondimensional variables $\bar{t}=\omega_p t$, $\bar{s}=\omega_p s$, $\bar{x}=x\omega_p/\wbbeta_x$,
$\bar{y}=y\omega_p/\wbbeta_y$, $\bar{z}=z\omega_p/\wbbeta_z$ and $\bar{r}=\sqrt{\bar{x}^2+\bar{y}^2+\bar{z}^2}$.
The shielding charge as a function of $\bar{r}$ and time is
\begin{equation}
\eta^1(\bar{r},t) = \frac{4Z}{\pi} \int\limits_0^{\bar{t}}
\frac{\bar{s}\sin(\bar{s}) \bar{r}^2}{\left[\bar{s}^2 + \bar{r}^2\right]^2} d\bar{s}
\end{equation}
Figure~\ref{fig:13} shows the shielding charge a distance $\bar{r}$ from a gold ion ($Z=79$) in intervals of a quarter plasma period.
Note that $\eta^1(\bar{r},t)$ is always positive for $0\le \bar{t}\le \pi$, or $t$ less than half a plasma period,
because the integrand is always positive.
This cannot happen in a finite domain, because in a finite domain the total charge perturbation must be zero.
To make up for the shielding near the ion, the charge perturbation
must be negative far from the ion (negative indicates a lack of electrons).

If the total shielding charge is $Q(\bar{t})$ then
\begin{equation}
Q = \frac{4Z}{\pi}\int\limits_0^\infty \int\limits_0^{\bar{t}}
\frac{\bar{s}\sin(\bar{s}) \bar{r}^2}{\left[\bar{s}^2 + \bar{r}^2\right]^2} d\bar{s} d\bar{r}
\end{equation}
Now we reverse the order of integration, and use the fact that
\begin{equation}
\int_0^\infty \frac{\bar{r}^2 d\bar{r}}{\left[\bar{s}^2 + \bar{r}^2\right]^2} = \frac{\pi}{4\bar{s}}
\end{equation}
to obtain the shielding charge as a function of time,
\begin{equation}
Q = Z(1 - \cos \bar{t}) = Z(1 - \cos \omega_p t)
\label{eq:scharge}
\end{equation}
The total shielding charge reaches a maximum of $2Z$ at a time of half a plasma period, $\bar{t}=\pi$, or $t = \pi/\omega_p$.
At a full plasma period, the total shielding charge is zero!
This is not obvious from Figure~\ref{fig:13}, because of the long tail on the distribution.

Although (\ref{eq:scharge}) was derived for a stationary ion, it also holds for an ion moving at constant velocity.
To see this, we start from the integrated charge distribution (\ref{eq:exactsol2})
and integrate over the remaining spatial variable $x$.
We non-dimensionalize as before, including the ion velocity \mbox{$\bar{v}^i_x = v^i_x / \wbbeta_x$}.
\begin{eqnarray}
Q & = & \int_{-\infty}^\infty \eta^1(x,t) dx \\
 & = & \frac{Z}{\pi} \int_{-\infty}^{\infty} \int_0^{\bar{t}} \frac{\bar{s}\sin(\bar{s}) d\bar{s} d\bar{x}}{{\bar{s}}^2 + (\bar{x}+\bar{s} \bar{v}^i_x)^2}
\end{eqnarray}
Again we switch the order of integration, and the integral over $\bar{x}$ can be calculated exactly, giving
\begin{equation}
Q = Z\int_0^{\bar{t}} \sin(\bar{s}) d\bar{s} = Z(1 - \cos \omega_p t)
\end{equation}
as before.


\begin{thebibliography}{9}   
\bibitem{Qian94}
Q. Qian et. al., ``Nonlinear $\delta$f simulation studies of intense ion beam propogation through
an alternating-gradient quadrupole focusing field", Phys Plasm., 10.1063 (2997).
\bibitem{Rosenzweig}
J.B. Rosenzweig, ``Fundamentals of beam physics", Oxford University Press, 2003, p. 119.
\bibitem{NX06}
N. Xiang, J.T. Cary, D.C. Barnes, ``Low-noise electromagnetic $\delta$f particle-in-cell simulation of electron Bernstein waves",
Phys. Plasmas {\bf 13} 062111 (2006).
\bibitem{WB08}
G. Wang and M. Blaskiewicz, ``Dynamics of ion shielding in an anisotropic plasma", Phys Rev E 78, 026413 (2008).
\bibitem{Bell08}
G.I. Bell et. al., ``Simulating the dynamical friction force on ions due
to a briefly co-propagating electron beam", J. Comput. Phys., 227, 87148735 (2008).
\bibitem{Lit09}
V.N. Litvinenko and Y.S. Derbenev, ``Coherent electron cooling", Phys. Rev. Lett. 102, 114801 (2009).
\bibitem{Bell11}
G.I. Bell et. al., ``Vlasov and PIC simulations of a modulator section for 
coherent electron cooling", PAC 2011 Proceedings, MOP067.
\bibitem{Bell12}
G.I. Bell et. al., ``High-fidelity 3D modulator simulations of coherent electron cooling systems",
IPAC 2012 Proceedings, THEPPB002.
\bibitem{Schwartz13}
B.T. Schwartz et. al., ``Coherent electron cooling: status of single-pass simulations".
IPAC 2013 Proceedings, MOPWO071.
\end{thebibliography}
\end{document}